\documentclass[12pt,letterpaper,usenames,dvipsnames]{article}
\usepackage{jheppub}
\usepackage{amsmath,amssymb,amsfonts,amsthm,amscd}
\usepackage{bm} 
\usepackage{bbm} 
\usepackage{eufrak} 
\usepackage{mathrsfs} 
\usepackage{lscape} 
\usepackage{subcaption} 
\usepackage{cancel}
\usepackage{enumitem}
\usepackage{tcolorbox}
\usepackage{multirow} 
\usepackage{colortbl} 
\usepackage{array,hhline,arydshln,multirow}
\usepackage{floatrow}
\usepackage[export]{adjustbox}
\usepackage{rotating} 

\allowdisplaybreaks[2]  
\usepackage{pifont,dsfont}
\usepackage{bm} 

\usepackage{graphicx,tikz}
\usepackage[framemethod=TikZ]{mdframed} 
\usepackage{tikz-cd} 
\usepackage[all]{xy}

\usepackage{xcolor}

\definecolor{amaranth}{rgb}{0.9, 0.17, 0.31}
\definecolor{coolblack}{rgb}{0.0, 0.18, 0.39}
\definecolor{gold(web)(golden)}{rgb}{1.0, 0.84, 0.0}
\definecolor{deepcarmine}{rgb}{0.66, 0.13, 0.24}

\def\bM{\begin{matrix}}
\def\eM{\end{matrix}}
\newcommand{\bpm}{\begin{pmatrix}}
\newcommand{\epm}{\end{pmatrix}}
\newcommand{\bsm}{\begin{smallmatrix}}
\newcommand{\esm}{\end{smallmatrix}}
\newcommand{\bspm}{\left(\begin{smallmatrix}}
\newcommand{\espm}{\end{smallmatrix}\right)}
\newcommand{\beq}{\begin{equation}}
\newcommand{\eeq}{\end{equation}}
\def\bar{\overline}

\def\^{\wedge}

\def\cC{{\mathcal C}}

\def\cL{{\mathcal L}}

\def\cN{{\mathcal N}}

\def\R{\mathbbm{R}} 

\def\cS{{\mathcal S}}

\def\cT{{\mathcal T}}

\def\Z{\mathbbm{Z}}

\def\D{{\Delta}}

\def\k{{\varkappa}}
\def\l{{\lambda}}

\usepackage{tikz}
\usetikzlibrary{arrows,snakes,shapes.arrows,decorations.markings}
     \tikzset{>=triangle 90}
     \tikzstyle{bbc}=[draw,circle,fill=black,scale=.75]
     \tikzstyle{rc}=[circle,fill=red,scale=.6]
     \tikzstyle{wc}=[draw,circle,scale=.75]

\def\dgreen#1{{\color{ForestGreen}{#1}}}

\def\bar{\overline}

\def\^{\wedge}

\def\dim{{\rm dim}}

\def\D{{\Delta}}
\def\bD{{\boldsymbol{\Delta}}}

\def\k{{\varkappa}}
\def\l{{\lambda}}

\def\cC{{\mathcal C}}

\def\cL{{\mathcal L}}

\def\cN{{\mathcal N}}

\def\cS{{\mathcal S}}

\def\cT{{\mathcal T}}

\def\CC{\mathbb{C}}

\def\R{\mathbb{R}} 
 
\def\Z{\mathbb{Z}} 

\def\beq{\begin{equation}}
\def\eeq{\end{equation}}
\def\be{\begin{equation}}
\def\ee{\end{equation}}

\newcommand{\bpmat}{\begin{pmatrix}}
\newcommand{\epmat}{\end{pmatrix}}
\newcommand{\bsmat}{\begin{smallmatrix}}
\newcommand{\esmat}{\end{smallmatrix}}
\newfloatcommand{capbtabbox}{table}[][\FBwidth]

\theoremstyle{theorem}

\theoremstyle{definition}

\newtheorem{fact}{Fact}

\def\Dsl{\,\raise.15ex\hbox{/}\mkern-13.5mu D}
\def\dsl{\,\raise.25ex\hbox{/}\mkern-10.5mu \partial}

\title{The Characteristic Dimension \\ of Four-dimensional $\mathcal N=2$ SCFTs}
\author[1]{Sergio Cecotti}
\author[2,3]{Michele Del Zotto}
\author[4,5]{Mario Martone}
\author[2]{Robert Moscrop}

\affiliation[1]{SISSA, via Bonomea 265, I-34100 Trieste, Italy}
\affiliation[2]{Mathematics Institute, Uppsala University,
Box 480, SE-75106 Uppsala, Sweden}
\affiliation[3]{Department of Physics and Astronomy, Uppsala University,
Box 516, SE-75120 Uppsala, Sweden}
\affiliation[4]{C.~N.~Yang Institute for Theoretical Physics,  Stony Brook University,Stony Brook, NY 11794-3840, USA}
\affiliation[5]{Simons Center for Geometry and Physics, Stony Brook University, Stony Brook, NY 11794-3840, USA}
\emailAdd{cecotti@sissa.it, michele.delzotto@math.uu.se, mmartone@scgp.stonybrook.edu, robert.moscrop@math.uu.se}

\abstract{In this paper we introduce the characteristic dimension of a four dimensional $\cN=2$ superconformal field theory, which is an extraordinary simple invariant determined by the scaling dimensions of its Coulomb branch operators. 
We prove that only nine values of the characteristic dimension are allowed, $-\infty$, 1 ,6/5, 4/3, 3/2, 2, 3, 4, and 6, thus giving a new organizing principle to the vast landscape of 4d $\mathcal N=2$ SCFTs.  Whenever the characteristic dimension differs from 1 or 2, only very constrained special K\"ahler geometries 
(\textit{i.e.} isotrivial, diagonal and rigid) are compatible with the corresponding set of Coulomb branch dimensions and extremely special, maximally strongly coupled, BPS spectra are allowed for the theories which realize them. Our discussion applies to superconformal field theories of arbitrary rank, \emph{i.e.} with Coulomb branches of any complex dimension. Along the way, we predict the existence of new $\cN=3$ theories of rank two with non-trivial one-form symmetries.
}

\begin{document}
\maketitle 

\section{Introduction}

Superconformal algebras (SCAs) owe their existence to little miracles in representation theory, the exceptional isomorphisms between Lie algebras in low ranks. This allows to fully classify them, and, in particular, rule out the existence of superconformal field theories (SCFTs) in dimensions larger than six \cite{Nahm:1977tg}. Solely the properties of SCAs and their representations give very stringent constraints on the dynamics of SCFTs -- for instance in 6d the maximal superconformal algebra compatible with the existence of a stress tensor multiplet is the 6d (2,0) one \cite{Cordova:2016emh}. SCFTs themselves are indeed so constrained that a full classification might be possible, especially in cases with a high enough amount of supercharges. Most notably, six-dimensional (2,0) theories have an ADE classification, that has a geometric origin in the McKay correspondence and IIB geometric engineering \cite{Witten:1995zh}, and a field theory origin in the assumption that these systems are the UV completion of 5d $\mathcal N=2$ gauge theories \cite{Cordova:2015vwa}. Six-dimensional (1,0) SCFTs also admits a classification within F-theory \cite{Heckman:2013pva,DelZotto:2014hpa,Heckman:2015bfa,Bhardwaj:2019hhd} whose field theoretical counterpart is an interesting open question.\footnote{\ For 6d SCFTs with conventional matter, a field theoretical classification can be found in \cite{Bhardwaj:2013qia}.} 

\medskip

A wide array of results have been obtained recently about SCFTs in dimension five and four starting from such classifications, 
leading to the conjecture that all lower dimensional SCFTs have a six-dimensional origin of sort. This conjecture has been formulated in the context of 5d SCFTs where it was checked for theories of rank one and two \cite{Jefferson:2018irk}. The results of \cite{Closset:2021lhd} building upon the special arithmetic of flavor \cite{Caorsi:2018ahl} can be exploited to extend the conjecture to 4d $\cN=2$ theories as well. Indeed, the physical interpretation of the result of \cite{Caorsi:2018ahl} is that via toroidal twisted compactifications of the (1,0) E-string theory one can obtain all rank-one 4d $\cN=2$ SCFTs that have been classified independently exploiting the special geometry of rank-one Coulomb branches \cite{Argyres:2015ffa,Argyres:2015gha,Argyres:2016xua,Argyres:2016xmc,Caorsi:2019vex}. More recently a study of the higher dimensional origin of the currently known rank-two $\cN=2$ theories in 4d was performed \cite{Martone:2021drm} with the result that overwhelmingly the latter do descend from higher dimensional SCFTs.

\medskip

These results give strong motivations to probe such a conjecture further. With that objective in mind it is crucial to develop other classification schemes that are intrinsically independent from 6d constructions. The first step in such a program is to identify a class of examples that are constrained enough that alternative routes of classification are possible. The second step is to develop a classification scheme which is based solely on the properties of that class and has no further input from geometric engineering or other types of constructions. This is the main aim of this paper, which is the first in a series dedicated to making strives towards classifying 4d $\cN \geq 2$ SCFTs of arbitrary rank $r \geq 1$ and thus quantitatively addressing this question. We propose a new classification scheme, which is inspired by the Kodaira-Enriques classification of compact complex surfaces \cite{compactcomplex} and is based on the new notion of \textit{characteristic dimension of four dimensional $\mathcal N=2$ SCFTs.} 

\medskip

Since we presume that not all our readers are familiar with the works by Enriques and Kodaira, let us briefly review how that story works. Kodaira built on Enriques’ work, who had the crucial insight of introducing an equivalence relation – birational equivalence – emphasizing that the meaningful program was to classify complex surfaces modulo this equivalence. Kodaira then made the breakthrough of introducing a new numerical birational invariant for surfaces – the Kodaira dimension $\k$ – which takes only a finite set of values
\beq
\k\in\{-  \infty, 0, 1, 2\}
\eeq
Then he proceeded to classify birational classes of compact surfaces $S$ with a given Kodaira dimension $\k(S)$. For $\k \notin\{1,2\}$ his classification is pretty explicit and detailed, while for $\k = 1, 2$ it captures only the general features. The case $\k=- \infty$ is mostly boring, a part from the existence of the special family of non-algebraic surfaces of type VII, that are still poorly understood.

\medskip

Inspired by the above ideas, in this paper we identify a numerical invariant associated to each 4d $\mathcal N=2$ SCFT that we call its characteristic dimension, and we denote with the letter $\k$. We show that the characteristic dimension of a rank $r\geq 1$ SCFT can take only a finite set of eight positive rational values, in one-to-one correspondence with the allowed rank-one CB dimensions
\beq\label{eq:rankonedims}
\k \in \left\{1,\frac65,\frac43,\frac32,2,3,4,6\right\}
\eeq
This can be used to organize the space of $\cN\geq 2$ SCFTs in a way similar to the Kodaira-Enriques story. Ironically, also in this case it turns out that when our invariant has value that differs from 1 or 2 the classification is pretty explicit and detailed, while it becomes rather schematic otherwise. For the sake of analogy, then, we decided to assign a ninth value, $\k=-\infty$, to the class of theories without a CB.\footnote{\ In that context we know of many boring examples, free hypers or discrete gaugings thereof \cite{Closset:2020scj,Closset:2020afy}, and we still do not know whether a family of interacting $r=0$ SCFTs exists or not. If it does, in our classification scheme, its name ought to be a type VII.} Since in this paper we do not consider any examples belonging to this category, we will assume $\k > 0$ from here onwards for the sake of simplicity of exposition. We proceed with a brief schematic exposition of our

\medskip

\noindent\textbf{Main results}:
\begin{enumerate}
    \itemsep-.05cm 
    \item The characteristic dimension $\k(\mathcal T)$ of a rank $r \geq 1$ SCFT $\mathcal T$ has only eight allowed values in one-to-one correspondence with the allowed rank-one CB dimensions \eqref{eq:rankonedims}.
    \item For theories with a freely generated Coulomb branch (CB) of dimension $r\geq 1$, $\k(\mathcal T)$ is captured by a simple formula, depending solely on the scaling dimensions of its CB operators --- see the discussion around equation \eqref{Chardim} below.\footnote{\ There is a more general explicit formula which holds under the weaker condition that the Coulomb branch chiral ring $\mathscr{R}$ is finitely generated. Let $\lambda$ be the largest rational number such that the Hilbert series of $\mathscr{R}$ (as a ring graded by the $\mathbb{C}^\times$-weights) is a rational function of $t^\lambda$. Then $\varkappa$ is defined as the unique rational number in $[1,+\infty)$ such that $1/\varkappa\equiv 1/\lambda\bmod 1$.}
    \item Theories for which $\k(\mathcal T) \notin \{1,2\}$ are severely constrained, in particular
    \begin{enumerate}
        \itemsep-.05cm 
        \item $\mathcal T$ is \textit{maximally strongly-coupled} (mSC), meaning that all energy levels in the spectrum over a generic point along the CB are necessarily degenerate with mutually non-local BPS states.
        \item $\mathcal T$ is a \textit{rigid SCFT}, by which we mean that $\mathcal T$ does not have an $\cN=2$ conformal manifold.\footnote{\ It can happen that rigid SCFTs are nevertheless not isolated, as they could still admit an $\cN=1$ conformal manifold along the lines discussed in \cite{Razamat:2019vfd}.}
        \item The special K\"ahler geometry (SKG) associated to the theory's CB is necessarily \textit{isotrivial}, \textit{i.e.} the CB metric $\tau_{ij}$ is locally constant, that is, globally constant over the entire CB\footnote{ \ The name isotrivial follows from the fact that the fact that $\tau_{ij}$ is constant implies that the total space of the CB is an \emph{isotrivial} fibration, \emph{i.e.} all smooth fibers are isomorphic as Abelian varieties.} up to $Sp(2r,\Z)$ rotations of the electro-magnetic duality frame. We call $\cN = 2$ SCFTs with isotrivial CBs, \emph{isotrivial SCFTs}.
        \item discrete $\mathbb Z_n$ subgroup of the $U(1)_R$ symmetry is necessarily unbroken at all generic points of the CB, with enhancements at non-generic points. We call this $\Z_n$ symmetry the \emph{characteristic symmetry} of the SCFT and we have
        \beq n = \begin{cases} 6 & \text{if } \k = \frac65,6\\ 4 & \text{if } \k = \frac43,4 \\ 3 & \text{if } \k = \frac32,3 \end{cases}\eeq
        This in particular entails that, in a suitable electro-magnetic duality basis, at a generic point
        \beq\label{diago} \tau_{ij} = \zeta \delta_{ij} \qquad \text{with}\quad \zeta^n =1\eeq
        modulo a local $Sp(2r,\mathbb{Z})$ rotation.\footnote{ \ Well-known examples of theories in such class are for example the rank-$r$ theories engineered as worldvolume theories of $r$ D3 branes probing F-theory 7-branes singularities and/or S-folds \cite{Banks:1996nj,Douglas:1996js,Sen:1996vd,Dasgupta:1996ij,Garcia-Etxebarria:2015wns,Aharony:2016kai,Apruzzi:2020pmv,Argyres:2020wmq,Giacomelli:2020jel,Giacomelli:2020gee,Bourget:2020mez,Kimura:2020hgw,Heckman:2020svr} which, in particular, include all rank-1 theories.} We call SKGs satisfying \eqref{diago} \emph{diagonal} SKGs and the theories realizing them \emph{diagonal} SCFTs.
    \end{enumerate}
    \item \label{isotta} \textit{Closure upon stratification of isotrivial SCFTs.} CBs of $\mathcal N=2$ SCFTs of rank $r$ are stratified by singular loci supported in codimension $r-p$ where the dynamics is that of a rank $p$ SCFTs. The pattern of intersections of such singular loci give rise to a stratification of the CB \cite{Argyres:2020nrr,Argyres:2020wmq,Xie:2021hxd}. Isotrivial SCFTs form a class which is necessarily closed upon stratification, meaning that all the strata are isotrivial theories themselves.\footnote{\ In an upcoming work in this series we have shown that also the converse of this statement is true. If all the lower rank theories on the strata are isotrivial, then also the UV theory has to be isotrivial itself.}
    \item \textit{Exotic $\mathcal N=3$ theories.} The purpose of every good classification program is to discover exotics. In this paper, we give a small teaser for our readers by identifying two novel rank-two $\mathcal N=3$ exotic SCFTs -- see Table \ref{tab:ella} for their properties. We present a stringent consistency check on their existence building on the stratification properties of their CBs and Higgs branches in section \ref{sec:newexamples}.
\end{enumerate}

\begin{table}[]
    \centering
    \begin{tabular}{|c|c c c c c |}
    \hline
    $\mathcal T \phantom{\Big|}$ & $\boldsymbol{\Delta}$ & $12c$ & $24a$ & $\mathfrak{f}$ & $k_\mathfrak{f}$\\
    \hline
       $G_5 \phantom{\Big|}$  & \{6,12\} & 102 & 204 & $\mathfrak{u}(1) \times \mathfrak{u}(1)$ & $- $ \\
    $G_8 \phantom{\Big|}$  & \{8,12\} & 114 & 228 & $\mathfrak{u}(1)$ & $- $\\
    \hline
    \end{tabular}
    \caption{Data for the new exotic $\mathcal N=3$ SCFTs. 
    }
    \label{tab:ella}
\end{table}

\vspace{-.1cm}

\noindent \textbf{Remarks.}
\begin{itemize}
    \item[$i.)$] A consequence of the remark about characteristic symmetry is that two theories $\mathcal T$ and $\mathcal T'$ with $\k \notin \{1,2\}$ such that
        \beq
        r(\mathcal T) = r(\mathcal T') \quad \text{and} \quad \k(\mathcal T) = \k(\mathcal T') 
        \eeq
        must have CB special K\"ahler geometries that are necessarily locally isometric.\footnote{ \ That is, they admit isometric (finite) covers.}
    \item[$ii.)$] We stress here that there are isotrivial theories also among the theories that are not maximally strongly coupled, e.g. with $\k \in \{1,2\}$. In particular, all theories with $\cN\geq3$ supersymmetry are necessarily isotrivial, with $\cN=3$ theories being rigid and $\cN=4$ being non-rigid. Isotriviality is instead quite peculiar in the strictly $\cN=2$ context. All the examples we know so far of purely $\cN=2$ isotrivial SCFTs arise from D3 branes probing 7-branes in F-theory and their S-folds. For that class of models, $\varkappa$ is obviously identified with the deficit angle corresponding to the F-theory 7-brane stack, which gives an F-theory origin of its possible 8 values, as well as to the many special properties of the mSC models with $\varkappa \neq \{1,2\}$.
    \item[$iii.)$] \textit{Isotrivial classification problem.} The closure under stratification of which at point \ref{isotta} above gives a crucial insight: the class of 4d $\mathcal N=2$ SCFTs with an isotrivial CB geometry is ideal from the perspective of an inductive classification, with inductive parameter the rank of the theory. We are currently developing a classification of isotrivial SCFTs, which will appear in a future publication in this series.
    \item[$iv.)$] \textit{Physical interpretation of the closure under stratification.}  The closure under stratification of which at point \ref{isotta} has a natural physical explanation: the massless BPS spectrum of an arbitrary rank-$r$ isotrivial geometry should necessarily be described, anywhere on the CB, by an effective low-energy theory which is itself isotrivial.
\end{itemize}

\section{Characteristic dimension: the physical insight}\label{sec:heureka}

In this section we introduce the \emph{characteristic dimension} $\varkappa$ of a 4d $\cN=2$ SCFT,
explain its physical meaning, and describe some of its far-reaching implications. 

\subsection{A quick and dirty review of the Coulomb branch}

The bosonic part of the $\cN=2$ superconformal algebra contains, besides the usual $\mathfrak{so}(4,2)$ conformal generators
\be
P_\mu,\ K_\mu,\ M_{\mu\nu},\ D,
\ee
the R-symmetry algebra $\mathfrak{u}(2)_R=\mathfrak{u}(1)_R\oplus\mathfrak{su}(2)_R$
which acts on the $Q$-supercharges as the fundamental representation plus its complex conjugate. We write $R$ for the generator of $\mathfrak{u}(1)_R$ normalized to be $\pm \tfrac{1}{2}$ on the supercharges.

 A continuous family of supersymmetric quantum vacua of a 4d $\cN=2$ SCFT is called a punctured CB (pCB) iff the Abelian R-symmetry
 $\mathfrak{u}(1)_R$ is spontaneously broken in the vacua of the family. A pCB is called \emph{pure} if, in addition, the symmetry $\mathfrak{su}(2)_R$
 is unbroken; otherwise is called \emph{mixed} (or \emph{extended}). All our arguments and results apply without changes in full generality to any pCB whether pure or mixed. However, for ease of exposition 
 we shall take the pCB to be pure of positive dimension.
 
 A (pure) pCB is parametrized by a complex space of dimension $r$ ($r$ being the rank of the SCFT). The complex charge $R+iD$
 generates a holomorphic $\CC^\times$-action on the punctured CB which then is the union of open $\CC^\times$-orbits. 
 The (unpunctured) CB $\cC$ is the closure of the pCB with respect to the
 action of $\CC^\times$. The CB contains just one more point than the pCB -- the \emph{origin} $0$ of $\cC$ -- corresponding
  to the unique vacuum of the SCFT where the superconformal symmetry is \emph{not} spontaneously broken. $0$ is the unique
  closed $\CC^\times$-orbit in $\cC$, and hence it lays in the closure of all $\CC^\times$-orbits.
  
 Typically $\cC$ is a copy of $\CC^r$ with global coordinates the vevs of CB operators of the SCFT, $(u^1,\cdots, u^r)$ transforming with definite weights under $\CC^\times$
 \be\label{Caction}
 (u^1,\cdots,u^r)\to (\lambda^{\Delta_1} u_1,\cdots, \lambda^{\Delta_r} u_r),\qquad \lambda\in \CC^\times,
 \ee  
where the weights $\Delta_i$ are rational numbers $>1$\footnote{ \ If $\D=1$ the theory necessarily has a free sector.} called the \emph{CB dimensions.}

\subsection{Stratification versus the generic point}

The traditional approach to the IR physics of a $\cN=2$ SCFT is to distinguish between the ``boring'' generic points $u\in\cC$,
which represent vacua where the only light degrees of freedom are $r$ IR-free massless vector multiplets governed by the effective bosonic
Lagrangian\footnote{ \ The indices $i,j$ take the values $1,2,\cdots,r$. The coupling $\tau_{ij}(u)$ is a symmetric $r\times r$ complex matrix with positive-definite imaginary part which
depends holomorphically on the point $u\in \cC$.}
\be
\cL_\text{bos}=\mathrm{Im}(\tau_{ij}(u))da^i d\bar a^j + \frac{1}{4\pi}\Big(\tau_{ij}(u)\, F_+^iF_+^j+\text{h.c.}\Big),\qquad F_\pm^i \equiv\frac{1}{2}(F^i\pm i\tilde F^i),
\ee
and the ``interesting'' locus $\mathcal{S}_1\subset\cC$ which parametrize SUSY vacua
where additional degrees of freedom become massless. We stress that the coupling $\tau_{ij}(u)$ is determined
only modulo a $Sp(2r,\Z)$-rotation of the electro-magnetic duality frame, and hence the entries of $\tau_{ij}(u)$ are multi-valued holomorphic 
functions on $\cC$, as are the scalar fields $a^i$. The v.e.v.\! $a^i$ of the scalar fields, while not univalued functions on $\cC$, 
are still a valid set of local complex
 coordinates in some neighborhood of a ``boring'' point $u\in\cC$ (they are called \emph{special local coordinates}).
 
By supersymmetry, the ``interesting'' locus $\mathcal{S}_1$ (if not empty\footnote{ \ This happens only if the theory is free.}) is a closed analytic
subspace of pure complex codimension 1. With a little more work, one can show that $\mathcal{S}_1$ is an algebraic hypersurface in the affine space $\CC^r$
which is quasi-homogeneous for the $\CC^\times$-action \eqref{Caction}.\footnote{ \ For the class of SCFT we focus on, this will be show below
where the equation of $\mathcal{S}_1$ will be explicity written. }
In turn, $\mathcal{S}_1$ contains a co-dimension 2 locus $\mathcal{S}_2\subset \mathcal{S}_1$ where even more states get massless,
and then we have a codimension-3 locus $\mathcal{S}_3\subset \mathcal{S}_2$ and so on, until we reach $\mathcal{S}_r\equiv\{0\}$, the origin of $\cC$,
where the superconformal symmetry is fully restored and all UV degrees of freedom are ``light''. 
Each $\mathcal{S}_j$ is a union of closed $\CC^\times$-orbits.
The IR interactions of the degrees of freedom which are light
in each irreducible component $\mathcal{S}_{j,\alpha}$ of the locus $\mathcal{S}_j\setminus \mathcal{S}_{j+1}$ are described by some effective $\cN=2$ QFT of
 rank $j$. 
 
 The basic idea of the traditional approach is to analyze the IR dynamics along each special locus $\mathcal{S}_{j,\alpha}$,
 and then use this local information to determine the low-energy physics globally on $\cC$, and in particular the (multivalued) holomorphic function $\tau_{ij}(u)$,
 which yields the special K\"ahler metric 
 \be
ds^2= \mathrm{Im}(\tau_{ij}(u))da^i d\bar a^j
 \ee
on the CB $\cC$. $ds^2$ is the natural metric entering in the low-energy Lagrangian $\cL$. We refer to the geometry of $\cC$ endowed with this K\"ahler metric as \emph{special K\"ahler geometry} (SKG).\footnote{\ For a nice review about special K\"ahler geometries our readers can consult the manuscript \cite{Freed:1997dp}, as well as the textbook \cite{Cecotti:book}.} $ds^2$ is well-defined and smooth only in the open dense domain $\cC\setminus \mathcal{S}_1$ where it is not geodesically complete.\footnote{ \ That is, the ``singular'' locus $\mathcal{S}_1$ is at finite distance from a generic base point if the distances are measured with the metric $ds^2$. }
Informally we see $ds^2$ as a K\"ahler metric with ``singularities'' along $\mathcal{S}_1$. 
  
 This standard approach has some serious drawbacks.  We need to understand the IR physics of the several components $\mathcal{S}_{j,\alpha}$
  of $\mathcal{S}_j\setminus \mathcal{S}_{j+1}$ for all $j$'s, and the complexity of the massless sector grows quite rapidly with the
 codimension $j$ of the component.
 This may be easy enough when the local light degrees of freedom are asymptotically weakly-coupled, but 
 often their interactions are described by strongly-coupled (effective) SCFTs. In this case we may still proceed with the local analysis at $\mathcal{S}_{j,\alpha}$ 
 provided we have already encountered the relevant
 strongly-interacting SCFTs and we have a full understanding of their non-perturbative physics. 
It is quite possible that along some component $\mathcal{S}_{j,\alpha}$ we find a strongly-interacting SCFT that we don't know yet;
indeed this is inevitable for large $j$ since our knowledge of the zoo of higher-rank SCFTs is still rather poor. As a strategy to classify
SCFTs, the traditional approach is then rather dismal: we cannot even \emph{start} classifying rank-$r$ SCFT before \emph{completing} the
classification for all ranks $r^\prime< r$ in full detail. In addition, we have the major technical problem of being able to glue the several local geometries along the $\mathcal{S}_{j,\alpha}$'s
into a global SKG on the whole of $\cC$, a problem which looks (almost) hopeless for higher $r$'s.

In this note we switch gears and take the \emph{opposite} attitude. We fix once and for all a unique SUSY vacuum $|u\rangle$,
chosen to correspond to a \emph{very generic} point $u\in\cC$, and study the purportedly ``boring'' IR physics in the Hilbert space $H_u$
constructed by acting with smeared local operators on the fixed vacuum $|u\rangle$. We never leave the comfort zone of our ``good''
vacuum $|u\rangle$ whose IR physics is so well understood to be dubbed ``boring'' in the mainstream literature.  We claim that under certain favorable circumstances
(to be specified momentarily) we can recover the \emph{global} special K\"ahler geometry of $\cC$ from simple facts about the IR physics in the ``boring'' Hilbert space $H_u$.
Once we have determined the special K\"ahler geometry everywhere in $\cC$, we can (if we wish so) specialize the result
to sub-loci $\mathcal{S}_{j,\alpha}$ of arbitrary codimension $j$ thus reading the non-perturbative physics of the corresponding
effective rank-$j$ local SCFT.
Although our procedure works under special circumstances, the method applies in arbitrarily high rank $r$
and, most importantly, it works precisely when the traditional approach is less efficient, that is, whenever the local light degrees of freedom are inherently strongly-coupled everywhere in $\cC$. Thus our ``dual'' viewpoint complements the standard one in a nice and fruitful way.

\subsection{The unreasonable effectivness of the boring vacuum}

By the definition of the CB, the $U(1)_R$ symmetry is spontaneously broken in our ``boring'' vacuum $|u\rangle$. However, it needs not to be completely broken:
a discrete subgroup $\Z_n\subset U(1)_R$ may still be unbroken. We claim that
the global special K\"ahler geometry is \emph{essentially} determined by the datum of the order $n$ of the unbroken R-symmetry in the generic vacuum $|u\rangle$ provided this order is larger than 2. ``Essentially'' means the following: given $r$ and $n\geq3$, there is a unique $\CC^\times$-equivariant \emph{master} special K\"ahler manifold $M_{r,n}$
and a \emph{finite} set $\Xi_{r,n}$ of discrete subgroups of the isometry group of $M_{r,n}$ such that 
the list of special K\"ahler geometries describing SCFT with the given $r, n\geq3$ is
\be\label{skg1}
\Big\{M_{r,n}/G,\ G\in\Xi_{r,n}\Big\}.
\ee   
In particular, for given $r,n\geq3$ all manifolds are locally isometric. From the classification standpoint the small residual ambiguity
parametrized by the finite set $\Xi_{r,n}$ is very welcomed:  one gets at once \emph{all} possible
special K\"ahler geometries with the given invariants $r,n$.

Our pretense to be able to solve non-perturbatively the IR physics in \emph{all} vacua along $\cC$
(including the ones in the special loci $\mathcal{S}_{j,\alpha}$) 
just by knowing that the physics in the single ``boring'' vacuum $|u\rangle$ is invariant under 
a $\Z_{n\geq3}$ symmetry, may look \emph{unreasonable}  at first. However, it is not so; let us explain why.

Recall that only the $\cN=2$ Poincar\'e supersymmetry is realized linearly on the SUSY-invariant vacuum $|u\rangle$, the superconformal symmetry being broken. This entails that on $H_u$ it acts a complex conserved charge $Z$, the \emph{central charge} of the $\cN=2$ SUSY algebra. $Z$ is an additive conserved charge which commutes with all Noether (super-)currents;
this puts strong constraints on $Z$ as a quantum operator acting in $H_u$.
We recall the following well-known result omitting its derivation. First, $H_u$ decomposes into 
super-selected sectors of defined magnetic and electric charges $(m^j,e_i)$
\be\label{ooiqzza}
H_u=\bigoplus_{(m^j,e_i)\in\Lambda} H_{u,(m^j,e_i)},
\ee 
where $\Lambda$ is the rank-$2r$ lattice of the integrally quantized
magnetic and electric charges for the $r$ massless photons. Then $Z$ acts as multiplication by a constant
in each super-selected sector $H_{u,(m^j,e_i)}$. Together with conservation and additivity, this entails
that there are complex constants $(a^i, b_j)$ such that
\be
Z=a^i e_i+ b_j m^j\quad \text{in $H_u$.}
\ee
Of course, the constants $a^i$ depend on the choice of our generic vacuum $|u\rangle$;
indeed, locally around our ``boring'' vacuum, the functions $a^i(u)$ ($i=1,\cdots,r$) may be used as (local)
special coordinates in the sense of SKG. For a $\CC^\times$-symmetric SKG one has the basic relation
\be\label{spec0}
b_i=\tau_{ij}\,a^j
\ee
where $\tau_{ij}$ is the effective gauge coupling matrix in the vacuum $|u\rangle$.

The operator $Z$ needs not to commute with the symmetry generators which are broken in $|u\rangle$.
For $R$ we have
\be
e^{iR\alpha}\, Z\, e^{-iR\alpha}= e^{i\alpha}\,Z\qquad \alpha\in \R.
\ee  
Thus the statement that $\Z_n\subset U(1)_R$ is unbroken in $|u\rangle$ actually means 
the following:
\begin{fact}\label{ooopi}\textit{
If the ``boring'' Hilbert space $H_u$ contains a state with mass $m$, spin $s$, and flavor
quantum numbers $f_s$, which belongs to a SUSY supermultiplet of a given type
(\emph{e.g.}\! a general long supermultiplet, a vector supermultiplet, a hypermultiplet, \emph{etc.}),
and has a central charge $z\neq0$, then it contains another state with the same
mass, spin and flavor charges, contained in a SUSY supermultiplet of the same kind,
whose central charge is $Z=\zeta\, z$ where $\zeta=e^{2\pi i/n}$.}
\end{fact} 
\noindent In particular the spectrum of $Z$ in $H_u$ is invariant under multiplication by $\zeta$.  Let consider the various possibilities in turn. 
\begin{itemize}
    \item[$\bullet$] If $n=1$, $\zeta=1$ and the above statement is empty. Of course, we cannot solve non-perturbatively anything using only an empty statement.
    \item[$\bullet$] If $n=2$, $\zeta=-1$. There is always a state with the opposite central charge,
 say the PCT conjugate.\footnote{ \ The story is slightly more subtle because of the role of the 
 flavor charges, but the subtlety is inessential.} We cannot pretend to solve a QFT
 on the mere ground that it is consistent with PCT, otherwise we would solve \emph{all}
 QFTs in a single shot.
  \item Whenever $n\geq3$, our ``unreasonable'' claim states that  the SKG is (essentially) determined.
\end{itemize}
The claim can be proven in several ways:
in \cite{inprep} we show that it follows from the very existence of the Seiberg-Witten differential; here
we argue for this statement using arithmetics. Before going to the argument, let us explain \emph{physically} why having a generically unbroken $\Z_{n\geq3}$
R-symmetry is so constraining that fixes the SKG in (essentially) all details. The analysis below will give:
\begin{fact}\label{basicff} \textit{Assume $n\geq3$. Let $|\psi\rangle\in H_u$ be any eigenstate of $Z$ with non-zero eigenvalue, 
 and $e^{2\pi i R/n}|\psi\rangle$
its partner with the same quantum numbers. The two partner states are \emph{never}
mutually local.}
\end{fact}
Having mutually non-local states whose masses are of the same order is quite a constraining condition in QFT. For instance, it is strictly inconsistent in any weakly-coupled
or Lagrangian QFT.\footnote{ \ In a weakly-coupled Lagrangian QFT the monopoles must be hierarchically heavier than the electrically charged particles.
The ratio of monopole to electron masses is of order $1/g^2$ and blows up at weak coupling.} To have a consistent dynamics in presence of mutually non-local particles of roughly the same mass requires some subtle non-perturbative ``miracle''.  
In our situation \emph{most} states appear in mutually non-local, exactly degenerate $n$-tuples ($n\geq3$), and moreover this happens \emph{generically} (since $u$ is a generic vacuum).
Consistency of the dynamics in such an extreme situation is a formidable condition: 
\emph{unsurprisingly}, there is no consistent solution for almost all $n$'s, and when it exists the solution is essentially unique.
Physical consistency
 requires
the SCFT to be inherently strongly-coupled  for $n\geq3$, with \emph{no} asymptotic limit where some charged sub-sector gets weakly-coupled. In particular, 
the effective SCFTs
along the several special loci $\mathcal{S}_{j,\alpha}\subset\cC$ (of arbitrarily high codimension $j$) must be strongly-coupled everywhere on the respective locus. 
This also entails that for $n\geq3$ the SCFT is rigid, that is, its $\mathcal{N}=2$ conformal manifold reduces to a point.\footnote{ \ This prediction will be checked explicitly below.}

\subsection{Arithmetics of the characteristic symmetry}
Building upon the discussion above we see that precisely those cases that defy the naive perturbative intuition are more tractable from our perspective, as they result more constrained by the existence of an unbroken characteristic symmetry. 

\medskip

The vacuum $|u\rangle$ was chosen to be \emph{very} generic; it follows that  
the electric coefficients $\{a^1,\cdots, a^r\}$, that is the value of the special coordinates at the vacuum $u$, form a very generic $r$-tuple of complex numbers.
Since the complex numbers are uncountable and there are only countably-many algebraic numbers,
the $\{a^1,\cdots, a^r\}$ are \emph{linearly independent} over the algebraic closure $\overline{\mathbb{Q}}$
of $\mathbb{Q}$. 

\medskip

We are interested in the spectrum of the operator $Z$ in $H_u$ seen as a discrete
subset of $\CC$
\be\label{spec1}
\mathsf{spec}\,Z\equiv\Big\{a^i e_i+ b_j m^j,\ (m^j,e_i)\in\Lambda\Big\}\subset \CC.
\ee
By additivity of the central charge, $\mathsf{spec}\,Z$ is a $\Z$-submodule of $\CC$,
that is, 
\be
z_1,z_2\in \mathsf{spec}\,Z\quad\Rightarrow\quad m z_1+n z_2\in \mathsf{spec}\,Z\quad \text{for all }
m,n\in\Z.
\ee 
The linear independence of the $a^i$ over  $\overline{\mathbb{Q}}$ yields the inequality
\be
\dim_ {\overline{\mathbb{Q}}}\big(\mathsf{spec}\,Z\otimes_\Z \overline{\mathbb{Q}}\big)\geq r.
\ee

For $n\geq3$, $\zeta\equiv e^{2\pi i/n}$ is a non-real root of unity of order $n$, and
 the cyclotomic field $\mathbb{Q}(\zeta)$ has degree $\phi(n)$, where $\phi$ is the famous Euler totient function.
 Then, by additivity and invariance under multiplication by $\zeta$,
  if $z_1,z_2\in\mathsf{spec}\,Z$
 \be
 \big(m_1+\zeta m_2+\cdots +\zeta^{\phi(n)-1}\,m_{\phi(n)}\big)z_1+
 \big(n_1+\zeta n_2+\cdots +\zeta^{\phi(n)-1}\,n_{\phi(n)}\big)z_2\in \mathsf{spec}\,Z,
 \ee
 for all $m_a,n_b\in\Z$. In other words, the unbroken $\Z_n$ $R$-symmetry enhances
 $\mathsf{spec}\,Z$ from a $\Z$-submodule of $\CC$ to
 a $\Z[e^{2\pi i/n}]$-submodule. To simplify the analysis, we replace the module with the corresponding vector $\mathbb{Q}(\zeta)$-space, that is, 
   \be\label{spec4}
 \mathsf{spec}\,Z\otimes_\Z\mathbb{Q}=\sum_{i=1}^{2r/\phi(n)} \mathbb{Q}(\zeta)\,c_i \equiv\sum_{i=1}^{2r/\phi(n)}\big(m_{i,1}+\zeta m_{i,2}+\cdots +\zeta^{\phi(n)-1}\,m_{i,\phi(n)}\big)c_i\subset\CC
 \ee
 where now $m_{i,a}\in\mathbb{Q}$ and the $c_i$'s are suitable complex constants. Since $\zeta^k\in \overline{\mathbb{Q}}$, counting dimensions of vector spaces we get 
 \be
 r\leq \dim_ {\overline{\mathbb{Q}}}\big(\mathsf{spec}\,Z\otimes_\Z \overline{\mathbb{Q}}\big) \leq \frac{2r}{\phi(n)}.
 \ee
 Then $\phi(n)=1$ or $\phi(n)=2$. The first case corresponds to $n=1,2$ which are the two instances where we cannot say anything.
 The second case corresponds to $n=3,4,6$. In particular we have shown
 \begin{fact}\textit{
 The unbroken subgroup of $U(1)_R$ at a \emph{generic} point along the CB of a $\cN=2$ SCFT is $\Z_n$ with
 $n\in\{1,2,3,4,6\}$.}
 \end{fact}
 
\noindent Our claim states that whenever $n=3,4$ or $6$ we can essentially determine the global SKG in the sense of eqn.\eqref{skg1}.
 \medskip
 
The next observation is due to Gauss: for $n=3,4,6$ the ring $\mathbb{Z}[e^{2\pi i/n}]$  is a principal ideal domain.
Hence, a torsionless $\Z[\zeta]$-module is free; in down-to-earth terms, this means that we are entitled to be naive
 and simply 
take the $m_a$'s in eqn.\eqref{spec4} to be integers, that is,
 \be\label{spec5}
 \mathsf{spec}\,Z=\sum_{i=1}^{2r/\phi(n)} \mathbb{Z}[\zeta]\,c_i \equiv\sum_{i=1}^{2r/\phi(n)}\big(m_{i,1}+\zeta m_{i,2}+\cdots +\zeta^{\phi(n)-1}\,m_{i,\phi(n)}\big)c_i\subset\CC.
 \ee 
 This identifies the charge lattice $\Lambda\cong \Z^{2r}$ with the lattice $\Z[\zeta]^r$; the identification is intrinsic since $\exp(2\pi i R/n)$
 is a symmetry of the physics. 
 \medskip

 The charge lattice $\Lambda$ carries a skew-symmetric, non-degenerated, integral, bilinear pairing (called the \emph{polarization})
 \be
 \langle-,-\rangle\colon \Lambda\otimes\Lambda\to \Z
 \ee
 given by the Dirac electro-magnetic pairing of charges
 \be\label{spec6}
 \langle (m^j,e_i),(\tilde m^j,\tilde e_i)\rangle= e_i\tilde m^i- m^j\tilde e_i.
 \ee
  The polarization is called \emph{principal} iff it induces an isomorphism $\Lambda\cong\Lambda^\vee$. For simplicity we assume the polarization to be principal.\footnote{ \  
 With more work
 one can show that our conclusions hold \emph{mutatis mutandis} without this assumption. The main difference is that the theories without a principal polarization will have a non-trivial global structure, see e.g. \cite{DelZotto:2020esg,Closset:2020scj,Bhardwaj:2021zrt,Bhardwaj:2021pfz,Closset:2020afy,Hosseini:2021ged,Buican:2021xhs,Bhardwaj:2021wif} for some recent studies on the topic of 1-form symmetries of $\cN=2$ SCFTs. Moreover, this assumption is really strictly necessary only for \eqref{spec7}, the following equations are true independently from it.}
 Two states in $H_u$ are \emph{mutually local} iff the Dirac pairing of their electro-magnetic charges vanishes.
 
  The polarization must be consistent with the unbroken R-symmetry,
 hence it must be a polarization on the lattice $\Z[\zeta]^r$ which is consistent with its structure of $\Z[\zeta]$-module.
Such a polarization may always be written in the form
  \be\label{poqw12}
\langle\boldsymbol{v},\boldsymbol{w}\rangle=\pm \frac{1}{\zeta-\bar\zeta}\Big(H(\boldsymbol{v},\boldsymbol{w})-H(\boldsymbol{w},\boldsymbol{v})\Big),\quad \boldsymbol{v},\boldsymbol{w}\in \Z[\zeta]^r,
  \ee
where $H$ is a positive-definite Hermitian form 
\be
H\colon \overline{\Z[\zeta]^r}\times \Z[\zeta]^r\to \Z[\zeta],
\ee 
and the overall sign reflects the orientation of the charge space \cite{Caorsi:2018zsq}.
For $n=3,6$ or for $n=4$ and $r$ odd,  up to isomorphism 
there is a unique such Hermitian form which induces a principal polarization,
namely the diagonal one\footnote{\ These statements are not as innocent as they sound. Even if they may be roughly translated in the slogan \emph{``everything works as naively expected'',} their actual proofs are a piece of higher Number Theory, see \textbf{Proposition 6.4} in \cite{japjap}.}
\be\label{spec7}
H(\boldsymbol{v},\boldsymbol{w})=\boldsymbol{\bar v}^t\boldsymbol{w}\equiv\sum_{a=1}^r \bar v_a w_a.
\ee
For $n=4$ and $r$ even we have a second inequivalent possibility, which is isometric over $\mathbb{Q}$ to the diagonal one, so it leads to the same
local SKG. 

Let us now argue that \textbf{Fact \ref{basicff}} is true. We write $\boldsymbol{v}\in \Z[\zeta]^r\cong\Lambda$ for the
vector of electric and magnetic charge of the state; then
  \be\label{poqw123}
\pm\langle\boldsymbol{v},\zeta\boldsymbol{v}\rangle= \frac{1}{\zeta-\bar\zeta}\Big(H(\boldsymbol{v},\zeta\boldsymbol{v})-H(\zeta\boldsymbol{v},\boldsymbol{v})\Big)= H(\boldsymbol{v},\boldsymbol{v})\geq0,
  \ee 
with equality only in the sector $H_{u,(0,0)}$ of zero charge.

Comparing eqns.\eqref{spec0},\eqref{spec1},\eqref{spec5},\eqref{spec6} and \eqref{spec7} we get
\be
b_i\equiv \tau_{ij}\,a^j =\zeta\, a^i
\ee
so that, we have $\tau_{ij}=\zeta\,\delta_{ij}$ in an appropriate duality frame.
Our point $u\in\cC$ was chosen to be very generic, so that
the equation 
\be\label{poiqw12}
\tau_{ij}=\zeta\,\delta_{ij}
\ee
 holds everywhere in $\cC$ modulo a
local $Sp(2r,\Z)$ rotation. In particular, the gauge coupling is frozen to either
$i\,\delta_{ij}$ or $e^{2\pi i/3}\,\delta_{ij}$, and the SCFTs with $n\geq3$ are necessarily rigid mSC-theories, as expected
on physical grounds.

\medskip

The special K\"ahler metric locally reads
\be\label{skg6}
ds^2= \sin(2\pi/n)\, \sum_i d\bar a^i\, da^i
\ee
and is locally flat. Thus, as a singular K\"ahler space,
\be
\cC= \CC^r/\mathbb G
\ee
for some discrete subgroup $\mathbb G$ of the isometry group $U(r)\ltimes \CC^r$ of $\CC^r$.

\medskip

\subsection{Structure of isotrivial and diagonal special geometries}
Let us restate the above results in a nicer way.
 Recall that the special geometry of a $\cN=2$ SCFT is, in particular,
 a ($\mathbb{C}^\times$-isoinvariant) holomorphic integrable system, i.e.\!
 a holomorphic fibration over the CB
 \be\label{intsyst}
 \pi\colon X\to \cC\cong \mathbb{C}^r
 \ee 
 such that the total space $X$ is holomorphic-simplectic with Lagrangian fibers \cite{Seiberg:1994rs,Seiberg:1994aj,Donagi:1995cf} (see also \cite{Freed:1997dp}).
 The smooth fibers $X_u$ are polarized Abelian varieties of dimension $r$
whose period matrices $\tau_{ij}(u)$ are equal to the effective couplings we denoted by the same symbol.\footnote{ \ The group $Sp(2r,\Z)$ of duality frame rotations is then identified with the mapping class group of
the underling complex torus of $X_u$.}  

The special geometry is \emph{isotrivial} iff all smooth fibers are isomorphic (as polarized Abelian varieties) to a fixed Abelian variety $A$. From eqn.\eqref{poiqw12} we see that 
the geometry is \emph{diagonal} iff, in addition,
\be\label{Adiag}
A\cong \overbrace{\,E_\zeta\times E_\zeta\times \cdots\times E_\zeta\,}^{r\ \text{factors}}
\ee 
where $E_\zeta$ is the elliptic curve with period $\tau\equiv \zeta$. There are several isotrivial
special geometries which are not diagonal (they have necessarily $\varkappa=1,2$). An important example
of isotrivial geometry which is neither diagonal nor rigid is given by the $\cN=4$ SCFT with gauge group $G$: in this case the fixed Abelian variety $A$ is
\be
A=\mathbb{C}^r\big/[\Gamma_G\oplus \tau\,\Gamma_G^\vee]
\ee
with $\Gamma_G$ the weight lattice of $G$.

In the isotrivial case the integrable system \eqref{intsyst} takes the form
\be
(A\times \mathbb{C}^r)/\mathbb{G}\to \mathbb{C}^r/\mathbb{G}
\ee
where $A$ is the fixed Abelian fiber. The action of $\mathbb{G}$ on the first factor
factors through the automorphism group $\mathsf{Aut}(A)$ of $A$, that is we have a group
homomorphism
\be
\sigma\colon \mathbb{G}\to\mathsf{Aut}(A)
\ee 
and the total space of the integrable system is explicitly
\be
X\equiv A\times \mathbb{C}^r\big/\big[(a,x)\sim(\sigma(g)a,g x),\ \ g\in\mathbb{G}\big].
\ee
Thus an isotrivial rank-$r$ special geometry is specified by the following data:
a model Abelian variety $A$, a discrete group $\mathbb{G}$ of isometries
of $\mathbb{C}^r$, and a map $\mathbb{G}\to\mathsf{Aut}(A)$.
In the diagonal case $A$ should have the special form \eqref{Adiag}.

Classifying the isotrivial special geometry is reduced to listing the
allowed triples $(A,\mathbb{G},\sigma)$: there are only finitely many
for a given $r$. Details on the classification will be given in another paper of
the present series. We close this section with some preliminary
observation as a preparation to the following section.  

\subsection{Reflection groups and the discriminant}
The discrete group $\mathbb G$ does not act freely on $\CC^r$, and this leads to orbifold singularities of $ds^2$ along the several components of the special locus $\mathcal{S}_1$.
In the language of eqn.\eqref{skg1}, the master geometry $M_{r,n}$ is $\CC^r$ with the flat K\"ahler metric;
 to complete the classification of SKG with $n\geq3$
 it remains to determine the finite set $\Xi_{r,n}$ of allowed isometry subgroups $\mathbb G$.

The action of $\mathbb G$ must be compatible with the $\CC^\times$ one, so $\mathbb G$ leaves the origin fixed, and then $\mathbb G \subset U(r)$.
Being a discrete subset of the compact group $U(r)$, $\mathbb G$ is finite. We can assume that $\mathbb G$ acts irreducibly on $\CC^r$,
since otherwise the SKG decomposes into the product of lower rank SKGs.

If, in addition, we assume the CB $\cC$ to be a copy of $\CC^r$\footnote{ \ This includes the vast majority of SCFTs, for other cases see \cite{Bourget:2018ond,Argyres:2018wxu}.}, by the Shephard-Todd-Chevalley theorem \cite{Shephard:1954,Chevalley:1955}, $\mathbb G$ must be an irreducible (complex) reflection group of degree $r$. 
However, not all such reflection groups belong to $\Xi_{(n,r)}$.
The point is that the allowed singularities of the special K\"ahler metric
are of a restricted (mild) kind (since they should correspond to physically meaningful
SCFTs along $\mathcal{S}_1$) and only a few groups $\mathbb G$ produce singularities of the right type.

The coordinate ring of the quotient $\CC^r/\mathbb G$,
with $\mathbb G\subset U(r)$ a finite reflection group, is a free polynomial ring
in $u^1,\cdots, u^r$ where $u^i$ is an invariant homogeneous polynomial of degree $d_i$ in the coordinates $x^i$ of $\CC^r$.
The set $\{d_1,\cdots, d_r\}$ are called the degrees of the reflection group $\mathbb G$.
The Jacobian $J(x)$ of the polynomial map
$(x^1,\cdots, x^r)\to (u^1,\cdots,u^r)$ is a semi-invariant polynomial for $\mathbb G$, that is,
\be
J(gx)=\chi(g)\,J(x),\quad
 \text{for $g\in \mathbb G$}
 \ee
where $\chi(g)$ is a root of unity (i.e.\! $g\mapsto \chi(g)$ is a unitary character of $\mathbb G$). 
Then there is an integer $k$ such that $J(x)^k$ is truly $\mathbb G$-invariant; hence, a homogeneous polynomial
$D(u^1,\dots, u^r)$ in the basic invariants $u^i$ called the \emph{discriminant}.
The (non reduced) locus $\mathcal{S}_1$ is the algebraic hypersurface
\be
\mathcal{S}_1= \big\{D(u^1,\dots,u^r)=0\big\}\subset \CC^r,
\ee
The allowed groups $\mathbb G$ are determined by the condition that the singularities
along the several irreducible components of the discriminant $\mathcal{S}_1$ are of the appropriate kind.

\medskip

To put the question of the allowed $\mathbb G$'s in the proper perspective, it is convenient
to introduce an invariant which refines the order $n$ of the unbroken $U(1)_R$. This invariant is the characteristic dimension of the corresponding $\mathcal N=2$ SCFT.

\subsection{The characteristic dimension $\varkappa$} Since both the polarization and
$\tau_{ij}$ are diagonal, to simplify the notation we can consider the $r=1$ case, the conclusion being true for all $r$ (and $n\geq3$).

\medskip

Since $Z$ is constant in each sector $H_{u,(m,e)}$,   we may \emph{dually} rephrase
\textbf{Fact \ref{ooopi}}  by saying that the symmetry $e^{2\pi i R/n}\colon Z\to \zeta\,Z$ relates different super-selection sectors
\be
|\psi\rangle\in H_{u,(m,e)}\quad\Rightarrow\quad e^{2\pi i R/n}|\psi\rangle\in H_{u,(m^\prime,e^\prime)},
\ee
that is, the symmetry $\Z_n\subset U(1)_R$ acts by permutations of the Hilbert-space direct summands in  eqn.\eqref{ooiqzza}.
We wish to determine this action explicitly.

Consistency with charge quantization, additivity, and Dirac pairing yields 
\be
e^{2\pi i R/n}\colon \left(\begin{smallmatrix} m\\ e\end{smallmatrix}\right)\to  \left(\begin{smallmatrix} m^\prime\\ e^\prime\end{smallmatrix}\right)\equiv A  \left(\begin{smallmatrix} m\\ e\end{smallmatrix}\right)
\ee
where $A\in SL(2,\Z)$. Two $A$'s which are conjugate in $SL(2,\Z)$ are equivalent modulo a rotation of the duality frame,
so we are interested in the integral matrices $A$ modulo conjugacy in $SL(2,\Z)$. By construction $\zeta$ is an eigenvalue of $A$,
and since $A$ is real, the other eigenvalue is $\bar\zeta$. For $n\geq3$ $\zeta\neq\bar\zeta$, so $A$ is semi-simple (diagonalizable).
The conjugacy classes in $SL(2,\Z)$ are in one-to-one correspondence with Kodaira's exceptional fibers, and hence
the allowed $A$ correspond to the Kodaira fibers whose monodromy has order $n=3,4,6$, see table \ref{TableK}. In the last column of the table $\varkappa\equiv\Delta$ is just the CB dimension of the rank-1 SCFT with given monodromy matrix $A$. For each order $n$ we have two Kodaira fibers, the starred and the un-starred one. The starred fibers have $\varkappa=n$,
while the un-starred ones have $\varkappa=n/(n-1)$. The two possible $A$'s for a given $n$
correspond to the two signs in \eqref{poqw123}, that is, to the two orientations of the charge space.\footnote{ \ The two $A$'s with the same order $n$ are conjugate in $GL(2,\Z)$
by the Pauli matrix $\sigma_1\not\in SL(2,\Z)$ which flips the orientation of the charge 
space.
}

\begin{table}
\centering
\begin{tabular}{cccc}\hline\hline
Kodaira fiber & order $n$ & $A$ & $\varkappa$\\\hline
$II$ & $6$ & $\left(\begin{smallmatrix}1 & 1\\ -1 &0\end{smallmatrix}\right)$ & $6/5$\\
$II^*$ & $6$ & $\left(\begin{smallmatrix}0 & -1\\ 1 &1\end{smallmatrix}\right)$ & $6$\\
$III$ & $4$ & $\left(\begin{smallmatrix}0 & 1\\ -1 &0\end{smallmatrix}\right)$ & $4/3$\\
$III^*$ & $4$ & $\left(\begin{smallmatrix}0 & -1\\ 1 &0\end{smallmatrix}\right)$ & $4$\\
$IV$ & $3$ & $\left(\begin{smallmatrix}0 & 1\\ -1 &-1\end{smallmatrix}\right)$ & $3/2$\\
$IV^*$ & $3$ & $\left(\begin{smallmatrix}-1 & -1\\ 1 &0\end{smallmatrix}\right)$ & $4$\\\hline\hline
\end{tabular}
\caption{\label{TableK}Kodaira fibers with monodromy $A$ of order $n=6,4,3$.}
\end{table}

\medskip

$\varkappa$ is an invariant of the SCFT which \emph{refines} the previous invariant $n$, since it contains the information
on the (dual) action of the unbroken symmetry $e^{2\pi i R/n}$ on the charge sectors, 
in addition to the order $n$ of the unbroken R-symmetry.
\medskip 

For general $r$ the overall sign $\pm$ is common to all terms in \eqref{spec7}, so the dual action of $e^{2\pi i R/n}$ on the charge lattice
$\Lambda\cong\Z[\zeta]^r$ is given by the block-diagonal matrix $\mathrm{diag}(A,A,\cdots,A)$ where $A$ is one of the matrices in table
\ref{TableK}. Therefore  for all ranks $r$ we may replace the order $n$ by the finer invariant $\varkappa$ which specifies
the order $n$ as well as the action of $e^{2\pi i R/n}$ as a permutation of the Hilbert-space direct summands in eqn.\eqref{ooiqzza}.
\medskip

We give the explicit expression of $\varkappa$ for a theory $\cT$ with CB dimensions $\bD=(\Delta_1,\cdots,\Delta_r)\in \mathbb{Q}^r$ of the SCFT.
Write
\be
(\Delta_1,\cdots,\Delta_r)=\lambda(d_1,d_2,\cdots,d_r),\qquad \lambda\in\mathbb{Q}^\times,
\ee 
where the $d_i$'s are the unique integers with $\gcd(d_1,\dots, d_r)=1$ which represent the same point in projective space as
$(\Delta_1,\cdots, \Delta_r)$.
We set
\be\label{Chardim}
\k(\cT) = \frac{1}{\{\lambda^{-1}\}},
\ee
where, for $x\in \R$, $\{x\}$ is the unique real number equal to $x$ mod 1 and $0<\{x\}\leq 1$. The above discussion yields
\begin{fact}
\textit{For a 4d $\cN=2$ SCFT $\mathcal T$ the invariant $\varkappa(\cT)$ takes a value which is an allowed CB dimension in rank-1, that is,
\be
\k(\cT)\in\big\{1,6/5,4/3, 3/2, 2, 3,4,6\big\},
\ee
and the order $n$ of the generically unbroken R-symmetry is the order of $1/\varkappa$ in $\mathbb{Q}/\Z$. If $\varkappa\neq1,2$
(i.e.\! $n\neq1,2$) the SKG is essentially uniquely determined by $\varkappa$.}
\end{fact}

We call $\varkappa$ the \emph{characteristic dimension} of the SCFT. In rank-1 it coincides with the dimension in the usual sense.
All possible values of $\varkappa$ are realized by a SCFT (this already holds in rank-1).
\medskip

In the classification of the allowed finite groups $\mathbb G$ we need to distinguish the ones allowed for $\k(\cT)=n$ and the ones for
$\k(\cT)=n/(n-1)$ ($n\geq3$).

\section{A teaser from the classification of isotrivial SCFTs}\label{sec:newexamples}

As mentioned in passing, all rank-1 theories are isotrivial and since in this case $\k(\cT)\equiv \D$, where $\D$ is the CB scaling dimension of the coordinate parametrizing the one dimensional CB, it is extremely straightforward to identify which of the rank-1 theory is also rigid (of course at rank-1 all geometries are diagonal). Thus rank-2 is the first case where the question becomes more interesting. 

Recently one of the authors compiled a catalog of known rank-2 theories \cite[Table 1-3]{Martone:2021ixp} and it is straightforward to check how many entries in this list have $\k(\cT)\neq\{1,2\}$. We notice that all theories which satisfy this condition arise in F-theory as wordvolume theory on two D3 branes, with or without S-folds \cite{Aharony:2016kai}. The fact that these theories are diagonal, isotrivial and rigid is somewhat obvious from the F-theoretic perspective, but a natural question to ask is whether these are the only possible theories at rank-2 satisfying such conditions. 

We partially address this question here presenting examples of perfectly consistent diagonal, isotrivial and rigid SKG and which furthermore satisfy a series of non-trivial low-energy constraints. These geometries are associated to peculiar crystallographic complex reflection groups and the relevant properties of the corresponding SCFTs are reported in table \ref{tbl:n3thys}. The mapping between crystallographic complex reflection groups and consistent iso-trivial CBs is explained for example in \cite{Caorsi:2018ahl, Argyres:2019ngz}. All the examples that we discuss here have $\cN=3$ supersymmetry with a non-trivial one-form symmetry. This is manifested in the geometry by the fact that the abelian variety (constantly) fibered over the CB is not principally polarized. It would be extremely interesting to see if the theories we present below can be constracted in string theory.\footnote{\ It is tempting to conjecture that these models can be obtained from F-theory by exploiting a slight generalisation of the S-fold construction of \cite{Garcia-Etxebarria:2015wns,Aharony:2016kai} involving the full IIB duality group $GL^+(2,\mathbb Z)$, instead of just $SL(2,\mathbb Z)$. See e.g. \cite{Debray:2021vob} for a recent discussion which prompted this conjecture.}

\subsection{Consistency conditions from stratifications}
Let us start from analyzing these putative theories from the point of view advocated in \cite{Martone:2021ixp, Martone:2020nsy}. There, the stratification of the CB leads to highly non-trivial formulae for the central charges of theory derived from anomaly matching {\it \`a la} Shapere and Tachikawa \cite{Shapere:2008zf}. Furthermore, there is a rich interplay between the CB stratification and the strafication of the Higgs branch (HB) \cite{Bourget:2019aer} leading to stringent consistency conditions for our putative theories. We will briefly discuss this method before applying it to two examples. Since our examples are in rank-2, we will phrase the discussion under this assumption. We stress, however that the method can be generalised.

\medskip

First of all, given CB dimensions $\{\Delta_u, \Delta_v\}$ associated to CB coordinates $\{u,v\}$, we can determine the polynomial forms of the irreducible components of the discriminant locus $D$. Conformal invariance implies that such polynomial $P_i(u,v)$ should necessarily be quasi homogeneous and we will call its scaling dimension $\D^{\rm sing}_i\equiv \D(P_i(u,v))$. We are always allowed a stratum of the form
\be\label{knot}
    u^p+\l v^q=0,
\ee
where $p$ and $q$ are chosen such that $p\Delta_u = q\Delta_v$ and $\gcd(p,q)=1$ and $\l\in\CC^\times$. To this stratum we associate a scaling dimension $\D^{\rm sing}\equiv p\Delta_u = q\Delta_v$. Additionally, we are allowed the stratum $u=0$ (resp. $v=0$) only when $\Delta_v$ (resp. $\Delta_u$) is a scaling dimension allowed at rank-1, \emph{i.e.} $\D_{v/u}=\{\frac65,\frac 43,\frac 32,2,3,4,6\}$. In this case $\D^{\rm sing}=\D_u$ or $\D_v$ depending on the case. Henceforth we use the following nomenclature \cite{Argyres:2018zay}:
\begin{itemize}
    \item \emph{knotted strata} to refer to strata identified by polynomials of the form \eqref{knot}.
    
    \item \emph{unknotted strata} to refer to strata identified by either $u=0$ or $v=0$.
\end{itemize}

\medskip

\medskip

Along any strata, we must find rank-one theories $\mathfrak{T}_{P(u,v)}$ which describe the set of BPS states which become massless there. The $\mathfrak{T}_{P(u,v)}$ are constrained by the Kodaira type corresponding to each stratum. Isotriviality restricts these to be genuinely interacting rank-1 SCFTs, a table of which can be found in \cite{Martone:2021ixp}, for example. With this set-up, we can now use the {\it central charge formulae} \cite{Martone:2020nsy}:
\begin{gather} 
    12c = 4+h+\sum_{i=1}^{n} b^i\Delta^{\rm{sing}}_i, \\
    24a = 10+h+6\bigg(\sum_{i=1}^2 \Delta_i - r\bigg) +\sum_{i=1}^{n} b^i\Delta^{\rm{sing}}_i.
\end{gather}
Here $h$ is the dimension of the extended Coulomb branch (ECB) and $b_i\in\tfrac{1}{2}\mathbb{N}$ is a quantity associated to the rank-1 theory on the stratum given by
\be
    b^i=\frac{12c_i-2-h_i}{\Delta_i},
\ee
where $c_i$ and $h_i$ are the central charge and ECB dimension of the rank-1 theory on the $i$-th stratum. A particularly useful combination of these is given by
\be
    24(c-a)=h-2+\sum_{i=1}^{n} b^i\Delta^{\rm{sing}}_i-6\bigg(\sum_{i=1}^2 \Delta_i - r\bigg).
\ee
When the theory can be Higgsed down to free hypermultiplets, this coincides with the dimension of the HB $d_{\rm{HB}}$. Otherwise, there is an extra contribution to the HB dimension equal to the value of $24(a-c)$ of the residual SCFTs.

\medskip

The final step of this consistency check is to analyze the HB stratification. It is worth recalling the following principle, which further constrains the possible HB stratification.

\medskip
\noindent {\bf F-condition.} The simple factors of the flavor group of an SCFT act on the massless BPS spectrum which arise on at least one irreducible component of the complex codimension one singular locus.

\medskip

\noindent Taking this effect into account, we can then try to form a consistent HB stratification. If we find one, we claim that the SKG given is consistent from the field theory point of view and can posit that a corresponding rank-2 SCFT exists.

\subsection{The $G_5$ isotrivial SCFT}
\begin{table}[]
\begin{tabular}{|c|cc|cc|cc|cc|}
\hline
\multicolumn{9}{|c|}{\bf New $\mathcal{N}>2$ theories} \\ \hline \hline
$G$ & \multicolumn{1}{l}{$\Delta_{u,v}$} & $\k(\cT)$ & $\mathfrak{T}_v$ & $\mathfrak{T}_{\mathrm{knot.}}$ & $12c$ & $24a$ & $\mathfrak{f}_k$ & $k_{\mathfrak{f}}$ \\ \hline
$G_5$ & $\{6,12\}$ & $6$ & $\mathcal{S}_{\varnothing,3}^{(1)}$ & $\mathcal{S}_{\varnothing,3}^{(1)}$ & $102$ & $204$ & $\mathfrak{u}(1)\times\mathfrak{u}(1)$ & - \\
$G_8$ & $\{8,12\}$ & $4$ & - & $\mathcal{S}_{\varnothing,4}^{(1)}$ & $114$ & $228$ & $\mathfrak{u}(1)$ & - \\ \hline\hline
\end{tabular}
\caption{Central charges, Coulomb branch dimensions and stratification data for the putative $\mathcal{N}> 2$ theories. The theories $\mathcal{S}^{(1)}_{\varnothing,n}$ are rank-1 $\mathcal{S}$-fold theories defined in \cite{Giacomelli:2020jel}.}
\label{tbl:n3thys}
\end{table}

Consider the case of $\k(\cT)=6$ where the generic fiber over the CB is $S=E_{\rho}\times E_{\rho}$ with $\rho=e^{i\pi/3}$. It is known that the maximal automorphism group of $S$ is exactly the complex reflection group $G_5$ \cite{fujiki88}. As such, this complex reflection group defines the particular SKG of the system and we expect a corresponding $\mathcal{N}\geq 2$ theory associated to it. Furthermore, one can infer the stratification type and scaling dimensions of geometry by inspecting the cyclic reflection subgroups of $G_5$ (see \cite{lehrer2009unitary} for such data). In particular, using the results of \cite{hwang2009}, we find that the $G_5$ theory should have scaling dimensions $\{6, 12\}$ and two $IV^*$ strata. We can test the consistency of the putative theory using the conditions detailed in \cite{Martone:2021ixp}. Let us now follow that approach.

Since $\Delta_u=6$ is an allowed rank-1 dimension, the unknotted stratum $v=0$ can support a rank-1 theory. The other theory lies on the stratum given by
\begin{gather}
	u^2+v=0.
\end{gather}
Both strata have scaling dimension $12$, meaning the central charge formulae are simply
\be
    12c=4+h+12(b_1+b_2),\quad 24a=106+h+12(b_1+b_2),
\ee
where $h$ is the dimension of the ECB and $b_{1,2}\in \tfrac{1}{2}\mathbb{N}$ are parameters associated to the rank-$1$ theories living on the strata. Consequentially, the higgsable part of the Higgs branch has dimension
\be
    24(c-a)=h+12(b_1+b_2)-98.
\ee
Anticipating a possible $\mathcal{N}=3$ theory we can place a copy of $\cS_{\varnothing,3}^{(1)}$ on each stratum, consequentially taking $b_1=b_2=4$, and set $h=2$. Doing so immediately gives us the data in table \ref{tbl:n3thys} and that $a=c$. This property is required by $\mathcal{N}\geq 3$ SUSY, validating our suspicions of enhanced supersymmetry. We can summarize this stratification in the following Hasse diagram.
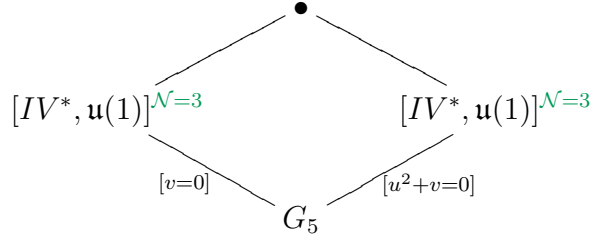
\begin{gather*}
	\xymatrix{ & \bullet \\
	[IV^*, \mathfrak{u}(1)]^{\dgreen{\mathcal{N}=3}} \ar@{-}@[blue][ur]& & [IV^*, \mathfrak{u}(1)]^{\dgreen{\mathcal{N}=3}} \ar@{-}@[blue][ul]\\
	& G_5 \ar@{-}@[blue][ul]^{[v=0]}\ar@{-}@[blue][ur]_{[u^2+v=0]}
	}
\end{gather*}
The remaining step is to check that using other rank-$1$ theories arising from $IV^*$ strata don't give rise to consistent CB and HB stratifications. This is easily done and leaves us with the stratification above.
\subsection{The $G_8$ isotrivial SCFT}
Similarly to the previous case, $G_8$ is known to be the maximal automorphism group of $E_i \times E_i$ \cite{fujiki88}. The process is much the same as before but with a minor caveat. 

Proceeding as before, we inspect the cyclic reflection subgroups of $G_8$ to extract the stratification type of the geometry. We see that there is conjugacy class consisting of $6$ order $4$ cyclic subgroups in addition to another conjugacy class with $6$ order $2$ subgroups. Again comparing with \cite{hwang2009}, we see that there is a $III^*$ stratum. However, inspecting the order 2 subgroups further shows that these arise simply as the index $2$ subgroups of the order $4$ cyclic groups. Therefore, the orbit of the $(+1)$-eigenspace of the generator coincides with those of the order $4$ groups and we obtain the same invariant and no extra information. We conclude that there is only a single $III^*$ stratum involved and nothing else. Additionally, the degrees of the invariants are $\{8, 12\}$, which correspond to the scaling dimensions.

Since neither of the scaling dimensions are of rank-1 type, the only allowable stratum is given by
\begin{gather}
    u^3+v^2=0,
\end{gather}
which has scaling dimension $24$. The central charge formulae take the form
\begin{gather}
    12c=4+h+24b,\quad 24a= 118+h +24b,
\end{gather}
which gives the higgsable HB dimension
\begin{gather}
    24(c-a)=h+24b-110.
\end{gather}
Anticipating an $\mathcal{N}=3$ theory, we take $h=2$ and set $24(c-a)$ to zero. Solving for $b$ tells us that $b=\tfrac{9}{2}$, indicating that the rank-$1$ theory supported on the $III^*$ stratum is the $\mathcal{N}=3$ $\mathcal{S}_{\varnothing, 4}^{(1)}$ theory. This gives us the linear Hasse diagram below.
\begin{gather*}
    \xymatrix{\bullet \ar@{-}@[blue][d] \\
               \quad [III^*, \mathfrak{u}(1)\rtimes \mathbb{Z}_2]^{\dgreen{\mathcal{N}=3}} \ar@{-}@[blue][d]^{[u^3+v^2=0]} \\
                G_8}
\end{gather*}
Checking the other possible rank-$1$ theories arising from a $III^*$ singularity shows that this is the only reasonable choice of stratification.

\acknowledgments The work of MDZ and RM has received funding from the European Research Council (ERC) under the European Union’s Horizon 2020 research and innovation programme (grant agreement No. 851931). MM is supported by NSF grants PHY-1151392 and PHY-1620610.

\bibliographystyle{JHEP}

\end{document}